\documentclass{llncs}
\usepackage{rotating}
\usepackage{amsmath}
\usepackage{amsfonts}
\usepackage{amssymb}
\usepackage{epic,eepic}

\def\BW{\put(15,5){\circle{3}}}
\def\CW{\put(25,5){\circle{3}}}
\def\DW{\put(5,15){\circle{3}}}
\def\EW{\put(15,15){\circle{3}}}
\def\FW{\put(25,15){\circle{3}}}

\def\AS{\put(5,5){\circle*{3}}\put(5,25){\circle{6}}}

\def\DS{\put(5,15){\circle*{3}}}

\def\FS{\put(25,15){\circle*{3}}}
\def\GS{\put(5,25){\circle*{3}}}

\def\IS{\put(25,25){\circle*{3}}}
\newcommand{\WZ}[1]{{\unitlength1pt\begin{picture}(30,32)(0,-2)
\path(0,0)(30,0)(30,30)(0,30)(0,0)(10,0)(10,30)(20,30)(20,0)(30,0)%
(30,10)(30,10)(0,10)(0,20)(30,20)\path(0,-2)(30,-2)#1\end{picture}}}
\newcommand{\SZ}[1]{{\unitlength1pt\begin{picture}(30,32)
\path(0,0)(30,0)(30,30)(0,30)(0,0)(10,0)(10,30)(20,30)(20,0)(30,0)%
(30,10)(30,10)(0,10)(0,20)(30,20)\path(0,32)(30,32)#1\end{picture}}}

\begin{document}

\pagestyle{headings}  

\mainmatter
\title{Attribute Exploration of Discrete Temporal Transitions}
\author{Johannes Wollbold}
\institute{Leibniz Institute for Natural Product
Research and Infection Biology - Hans-Kn\"oll-Institute (HKI)\\
Department of Molecular and Applied Microbiology / Systems
Biology Group\\
Jena, Germany\\
\email{jwollbold@gmx.de}} \maketitle

\begin{abstract}
Discrete temporal transitions occur in a variety of domains, but this work is mainly motivated by
applications in molecular biology: explaining and analyzing observed
transcriptome and proteome time series by literature and database knowledge. The
starting point of a formal concept analysis model is presented. The objects of a
formal context are states of the interesting entities, and the attributes are
the variable properties defining the current state (e.g. observed presence or
absence of proteins). Temporal transitions assign a relation to the objects,
defined by deterministic or non-deterministic transition rules between sets of
pre- and postconditions. This relation can be generalized to its transitive
closure, i.e. states are related if one results from the other by a transition
sequence of arbitrary length. The focus of the work is the adaptation of
the attribute exploration algorithm to such a relational context, so that
questions concerning temporal dependencies can be asked during the exploration
process and be answered from the computed stem base. Results are given for the
abstract example of a game and a small gene regulatory network relevant to a
biomedical question.
\end{abstract}

\section{Introduction}
Discrete temporal
transitions occur in a variety of domains: control of engineering processes or
roboters, flow of computer programs, a piece of music, games, etc. We are mainly
interested in
biological applications, but we develop a formal structure as widely usable as possible.

The practical aim is to explain experimental time series in molecular biology or to hypothesize
about temporal developments, especially in the context of gene expression regulation. Its first
step is transcription, i.e. the synthesis of mRNA from a DNA sequence coding for a gene.
Concentrations of mRNA for all genes of a cell culture (transcriptome analysis) can be measured by
the rather new technique of microarrays (RNA binds to matching fragments of DNA or RNA fixed on a
chip). The second step of gene expression is the translation of the mRNA into multiple identical
proteins by ribosomes. Since the mRNA concentrations are only weakly correlated to the respective
protein concentrations, it is recommended to also measure the latter, i.e. to perform proteome
analysis. However, it is unfavourable that weakly expressed proteins remain undetectable. By
complex - activating or inactivating - interactions of proteins within or between cells (signaling
pathways), a special class of proteins can be activated and - if necessary - transported to the
cell nucleus. Those transcription factors again regulate the expression of a sometimes large set of
genes.

\begin{figure}\label{network}
\vspace{0cm} \centering
\includegraphics[width=6cm]{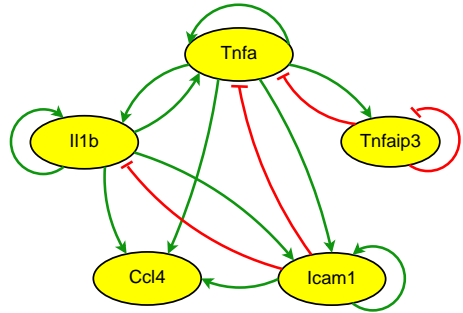}
\caption[]{\footnotesize Gene regulatory network. $\rightarrow$ upregulation, $\dashv$
downregulation. The information was obtained from the text mining software PathwayStudio
[www.ariadnegenomics.com] and the manually curated protein interaction database Transpath
[www.biobase.de].}
\end{figure}

At a more global level, such cycles are described as gene regulatory networks (Figure
\ref{network}). One abstracts from biochemical activation processes of proteins; only the mRNA or
protein level is considered as the main influencing factor. The indirect interactions between genes
are positive (upregulation of expression) or negative (downregulation). Regulatory networks may be
constructed based on knowledge available by manual or automatic (text mining) literature search and
in biological databases.

The network determines the possible transitions between properties of gene products (mRNA or
protein levels); as a first approximation they can be either present or absent. In the following we
translate similar situations into the language of formal concept analysis (FCA), so that attribute
exploration \cite[85ff.]{GW99} can be applied. During this interactive algorithm, an expert is
asked about the general validity of basic implications $A \rightarrow B$ between the attributes of
a given formal context $(G,M,I)$. An implication has the meaning: "If an object $g \in G$ has all
attributes $a \in A \subseteq M$, then it has also all attributes $b \in B \subseteq M$." If the
expert denies, he must provide a counterexample, i.e. a new object of the context. If he accepts,
the implication is added to the stem base (Duquenne-Guigues base) of the context. At the end, all
implications valid in the possibly enlarged context can be derived from the minimal set of rules
contained in the stem base. Those are identical to the implications valid in the explored domain
according to the knowledge available to the expert.

The present work is based on a FCA modeling of temporal transitions in \cite{Rud01}. The biological
application is influenced by computation tree logic \cite{Cha04}, Boolean networks \cite{Kau93} and
qualitative reasoning \cite{KGC05}. Temporal concept analysis as developed by K.E. Wolff
\cite{Wol05} is more directed toward a description of temporal concepts than toward temporal logic.
In future work, we shall investigate existing analogies and take advantage of them.

\section{Methods - Basic Definitions}
We start with two sets:
\begin{itemize}
\item The universe $E$. The elements of $E$ will be called \textbf{entities}.
They represent the objects of the world which we are interested in.
\item The set $F$ (\textbf{fluents}) denotes changing properties of
the entities.
\end{itemize}
A state of the universe is characterized by a unique value in $F$ taken by every
$e\in E$; states
with the same attribute values are identified.\footnote{They can of course be differentiated by
introducing a new attribute, e.g. "time interval".\\
The definition of a relational context $((G,R),M,I)$ developed below corresponds to a labeled
transition system with attributes, in the sense of \cite[Definition 1]{Rud01}.
It has a single action
''update'' or ''switch'' and is trivially attribute defined
\cite[Definition 2]{Rud01}.} Therefore a
state can be defined as a map $\varphi: E \rightarrow F$. If the state is not
completely known,
$\varphi$ is a partial map. To explore static features of states, the following formal context is
defined as a special case of a many-valued context \cite[36ff.]{GW99}. An
example of an attribute exploration of a state context (defined as a
single-valued
context and with a slightly more general notion of a state) is given in \cite[4.1.]{Rud01}.
\begin{definition}
Given two sets $E$ (entities) and $F$ (fluents),
a \textbf{state context} is a many-valued context $(G,E,F,J)$
with $G\subseteq \{\varphi: E \rightarrow F\}$; its relation
$J$ is given as $(\varphi,e,f)\in J
\Leftrightarrow \varphi(e) = f$, for all $\varphi \in G, e \in E$ and $f \in
F$.
\end{definition}
The class of these contexts is well defined; since $\varphi$ is a map, the
property of a
many-valued context is fulfilled: $(\varphi,e,f_1)\in J \wedge
(\varphi,e,f_2)\in J \Rightarrow f_1 = f_2.$ If a many-valued attribute is
regarded as a partial map from $G$ into $F$, one can also
write $e(\varphi)=f$.

For each attribute $e\in E$, a scale can be defined, i.e. a one-valued context $S_e:=(G_e,M_e,J_e)$
with $e(G)\subseteq G_e$. Thus by plain scaling we derive from $(G,E,F,J)$ the context $(G,M,I)$
with
\begin{equation}
M:=\bigcup_{e\in E} e \times M_e,\text{ and}
\end{equation}
\begin{equation}
\varphi I (e,m_e) : \Leftrightarrow e(\varphi) = f \text{ and }fI_e m_e.\
\end{equation}
If $\forall e \in E: M_e \subseteq F$, we get $M \subseteq E \times F$. This is the case e.g. for
nominal, ordinal and dichotomic scales. For nominal and dichotomic scales, the relation $I$ simply
is defined by $\varphi\, I\, (e,f) :\Leftrightarrow \varphi(e) = f$;
the following text is based on this relation.

Now we need a supplementary structure: a \textbf{relation} $R \subseteq G \times G$ indicates
temporal transitions between the states. A deterministic relation may be given by a family of
elementary \textbf{transition rules}: preconditions / postconditions $(V_k,N_k)_{k\in K}$, $V_k,
N_k \subseteq M$, so that
\begin{equation}\label{R}
 (\varphi_0,\varphi_1)\in R \Leftrightarrow \forall k \in K: V_k \subseteq
\varphi_0' \Rightarrow N_k \subseteq \varphi_1'.
\end{equation}

In the non-deterministic case (e.g. for a game), different postconditions are possible. There is a
class of families $\{(V_k,N_k^l)_{k\in K}|\: l\in L_k \text{ for all }k \in K\}$, and
\begin{equation}
\forall k \in K: V_k \subseteq \varphi_0' \Rightarrow \exists l \in L_k:
N_k^l \subseteq \varphi_1'
\end{equation}

The relational context $((G,R),M,I)$ can be represented by a binary power
context family. Here we
prefer the equivalent context, analoguous to \cite[Definition 4]{Rud01}:
\begin{definition}\label{defTransCon}
Given a state context $(G,E,F,J)$ and a relation $R \subseteq G\times G$, a \textbf{transition
context} $\mathcal{K}$ is the context $(R, M\times\{0,1\}, \tilde{I})$,
$M\subset E\times F$,
with the
property
\begin{equation}\label{trContext}
\forall i \in \{0,1\}\!:\: (\varphi_0,\varphi_1) \tilde{I} (e,f,i) \Leftrightarrow \varphi_i(e)=f.
\end{equation}
\end{definition}

It appears promising to consider the transitive closure $t(R) = \bigcup_{n\in \mathbb{N}} R^n$,
i.e. $\varphi_0\,t(R)\, \varphi_1$ for any elements $\varphi_0$ and $\varphi_1$ of $G$, provided
there exist $\alpha_0, \alpha_1, ..., \alpha_n \in G$ with $\alpha_0=\varphi_0,
\alpha_n=\varphi_1$, and $\alpha_r R\alpha_{r+1}$ for all $0\leq r < n$. That means, the state
$\varphi_1$ emerges from $\varphi_0$ by some transition sequence of arbitrary length. So we get a
new \textbf{transitive context}
\begin{equation}
\mathcal{K}_t:=((G,t(R)),M,I)\, \hat{=}\, (t(R), M\times\{0,1\},
\tilde{\tilde{I}}).
\end{equation}
The relation $\tilde{\tilde{I}}$ is defined like ${\tilde{I}}$
in (\ref{trContext}).

Regarding this context, queries like the following are possible, for $A,B,C\subseteq M,$ $m\in M$
(compare \cite[37]{Cha04}, \cite[2020f.]{KGC05}). In a non-deterministic setting, the implications
(\ref{qnever}) and (\ref{qalways}) refer to all possible transition paths starting from a state
$\varphi_0$ with all attributes $b \in B$. According to computation tree logic \cite[33]{Cha04},
one could also ask if a path exists with the respective property. (\ref{eventually}) expresses that
in the future development of $\varphi_0$, there will be a state with attribute $m$ for at least one
path.
\begin{align}
\label{qnever}B \rightarrow \text{never}(m)\:&\Leftrightarrow\: (B\times\{0\})'
\cap (m,1)'
= \emptyset\\
\label{eventually}B \rightarrow \text{eventually}(m)\:&\Leftrightarrow\:
(B\times\{0\})'
\cap (m,1)' \neq \emptyset\\
\label{qalways}B \rightarrow \text{always}(m)\:&\Leftrightarrow\:
(B\times\{0\})'
\subseteq (m,1)'\\
\label{oscillation}\exists \text{ stable state or
oscillation}\:&\Leftrightarrow\: \exists B\subset
M: (B\times\{0\})' \cap (B\times\{1\})' \neq \emptyset
\end{align}\\[-9mm]
\begin{equation}
\begin{aligned}
\label{3point}
&\text{Given a (partial) initial state $A$, can the system}\\
&\text{reach the state $C$ while passing by another state $B$?}\\
\Leftrightarrow\: &(A\times\{0\})' \cap (B\times\{1\})' \neq \emptyset
\:\wedge\, (B\times\{0\})'
\cap (C\times\{1\})' \neq \emptyset
\end{aligned}
\end{equation}
Those queries can also be checked for contexts modified by omitting some transition rules. So one
can investigate, if certain interactions are necessary for specific state transitions.

The attribute exploration process has to be adapted, so that similar questions can be asked as
implications during the exploration and be answered from the computed stem base. The following
equivalences are straightforward:
\begin{align}
B \rightarrow \text{never}(m)\:&\Leftrightarrow\: B\times\{0\}\cup (m,1)
\rightarrow \bot\label{never}\\
B \rightarrow \text{always}(m)\:&\Leftrightarrow\: B\times\{0\} \rightarrow (m,1)\label{always}
\end{align}
A counterexample has to be introduced into the context, if the temporal property in question is in
contradiction to the data or to the desired behaviour of the system which is to be designed.

\section{Results - Two Examples}
In this section a state transition
$(\varphi_0, \varphi_1)$ is written as $(\varphi^{in}, \varphi^{out})$,
and attributes are noted as $m^{in}$ or $m^{out}$ instead of
$(m,0)$ or $(m,1)$.
\subsection{3-pawns-chess}
In order to get a widely applicable view on discrete state transitions, the abstract case of a
simple game is introduced. It resembles chess with only three pawns. The game is won when a pawn
reaches the opposite side or when the opponent is blocked from further moves. Below are listed all
states reachable from a state $\varphi^{in}_0$ (0. - two moves after the beginning), and the bar
marks the next player. The following transitions are possible:
\begin{center}
$(\varphi^{in}_0, \varphi^{out}_1)$,
$(\varphi^{in}_0, \varphi^{out}_2)$, $(\varphi^{in}_0, \varphi^{out}_3)$;\\
$(\varphi^{in}_1, \varphi^{out}_4)$;\\
$(\varphi^{in}_2, \varphi^{out}_5)$, $(\varphi^{in}_2, \varphi^{out}_6)$
(similar transitions are not listed);\\
$(\varphi^{in}_3, \varphi^{out}_7)$.
\end{center}
In states 4, 5 and 7, black wins, in 6 white.

\begin{minipage}[t]{30mm}
\begin{enumerate}\setcounter{enumi}{-1}
\item \WZ{\BW\CW\GS\DS\IS}\\
\item \SZ{\CW\DW\GS\IS}
\end{enumerate}
\end{minipage}
\begin{minipage}[t]{30mm}
\begin{enumerate}\setcounter{enumi}{1}
\item \SZ{\EW\CW\GS\DS\IS}\\
\item \SZ{\BW\FW\GS\DS\IS}
\end{enumerate}
\end{minipage}
\begin{minipage}[t]{30mm}
\begin{enumerate}\setcounter{enumi}{3}
\item \WZ{\CW\DW\GS\FS}\\ 
\item \WZ{\EW\CW\AS\GS\IS} 
\end{enumerate}
\end{minipage}
\begin{minipage}[t]{30mm}
\begin{enumerate}\setcounter{enumi}{5}
\item \WZ{\EW\CW\GS\DS\FS}\\
\item \WZ{\BW\FW\GS\AS\IS}
\end{enumerate}
\end{minipage}\\

Our basic sets are $E = (\{a,b,c\} \times \{1,2,3\})\: \cup$ \{move, win\}, $F$
= \{white, black\}.
$G$, the set of all possible states of the game, is a proper subset of
$\{\varphi: E \rightarrow
F\}$. Some examples of the attributes are a1.white, move.white or win.black. The
state context
$(G,E,F,J)$ is not complete, because in every situation there are at least 3
empty fields, and not
every state is a win-situation; there exist $e \in E$, so that the domain
$D$ of the corresponding map $e: D\subseteq G \rightarrow F$ is not equal to
$G$.

Starting from the context with the transitive relation for the states 0. to 7., the stem base was
computed.\footnote{This is equivalent to an attribute exploration, where the expert accepts all
implications.} Among others, the following of the 61 implications are of some interest ($\top$
denotes the empty set of preconditions, $\bot:=M$ provided $M'=\emptyset$):
\begin{itemize}
\item $\top \rightarrow$ a3.black$^{in}$, c3.black$^{in}$, a3.black$^{out}$: a3
is always
occupied by black, c3 always but in the last step.
\item b2.black$^{out}$ $\rightarrow \bot$: b2.black$^{out}$ characterizes
an impossible
game situation.
\item  a2.white$^{out}$, move.white$^{out}$ $\rightarrow$ c2.black$^{out}$,
win.black$^{out}$:
For white, this implication could be a warning not to move to a2.
\item a2.black$^{out}$,  move.white$^{out}$ $\rightarrow$ win.white$^{out}$:
This
confirms the tactic importance of a2.
\item  c3.black$^{out}$, move.white$^{out}$ $\rightarrow$ a1.black$^{out}$,
win.black$^{out}$:
another winning condition.
\end{itemize}

\subsection{Gene regulatory networks}
We want to provide a temporal semantics for gene regulatory events, e.g.
''gene1 upregulates the
expression of gene2''. So the entities $E$ are the interesting
genes, and the fluents $F$ = \{abs, pres\} = \{-,+\} are mRNA or protein
levels.

In this section, the biological application of the present approach is explained by the example of
the 5 gene network of Figure \ref{network}. We confine ourselves to a single measured time series
of mRNA concentrations. It is part of ongoing biomedical research directed toward the understanding
of complex molecular interactions relevant for the pathogenensis and therapy of rheumatoid
arthritis (RA). This disease putatively has autoimmune causes, and it is recognized that proteins
like Tnf$\alpha$ and Il1$\beta$ - responsible for intercellular communication - have a major
stimulating influence on the inflammatory process \cite{Glo06}. Therefore fibroblasts (particular
cells of the joint) from RA patients were stimulated with Tnf$\alpha$, and their expression was
monitored by Affymetrix U133 Plus 2.0 microarrays before and 1, 2, 4 and 12 hours after
stimulation. mRNA levels were grouped into the two classes absent and present.\footnote{For larger
examples and datasets, a formal method will be selected, like the present/absent call of the gene
expression chip, cluster analysis or minimization of intra group variance.} One resulting time
course is shown in Table \ref{obsCon} as a transition context $\mathcal{K}^{obs}$ according to
Definition \ref{defTransCon}.

Now a corresponding knowledge based context will be developed. State transitions are computed
according to (\ref{R}): all rules of one family are applied with preconditions matching the
attributes of the input state $\varphi^{in}$. The type of rules valid for particular genes is
determined by the regulatory network (Figure \ref{network}). Table \ref{transRules} lists some
basic rule types; they are sufficient to compute the 2-gene transition context of Table
\ref{transCon}.
\begin{table}\caption{\footnotesize Observed transition context
$\mathcal{K}^{obs}$.}\label{obsCon}\centering
\begin{tabular}{|c|ccccc|ccccc|}
\hline\textbf{Transition}
&\begin{sideways}\textbf{Tnf$\alpha^{in}$}\end{sideways}
&\begin{sideways}\textbf{Tnfaip3$^{in}$}\end{sideways}
&\begin{sideways}\textbf{Icam1$^{in}$}\end{sideways}
&\begin{sideways}\textbf{Ccl4$^{in}$}\end{sideways}
&\begin{sideways}\textbf{Il1$\beta^{in}$}\end{sideways}
&\begin{sideways}\textbf{Tnf$\alpha^{out}$}\end{sideways}
&\begin{sideways}\textbf{Tnfaip3$^{out}$}\end{sideways}
&\begin{sideways}\textbf{Icam1$^{out}$}\end{sideways}
&\begin{sideways}\textbf{Ccl4$^{out}$}\end{sideways}
&\begin{sideways}\textbf{Il1$\beta^{out}$}\end{sideways}\\
\hline
$(\varphi_0^{in}, \varphi_1^{out})$ &+&-&-&-&-&+&+&+&+&-\\
$(\varphi_1^{in}, \varphi_2^{out})$ &+&+&+&+&-&+&+&+&+&+\\
$(\varphi_2^{in}, \varphi_3^{out})$ &+&+&+&+&+&+&+&+&+&-\\
$(\varphi_3^{in}, \varphi_4^{out})$ &+&+&+&+&-&-&+&+&-&+\\
\hline
\end{tabular}
\end{table}
\begin{table}\caption{\footnotesize Transition rules (simplified
notation).}\centering\label{transRules}
\begin{tabular}{lll}
\hline\noalign{\smallskip} \textbf{Nr.} &\textbf{Meaning} &\textbf{Rule}\\
\noalign{\smallskip} \hline \noalign{\smallskip}
1 &Upregulation &gene1.pres $\rightarrow$ gene2.pres \\
2a &Downregulation &gene1.pres, gene2.pres $\rightarrow$ gene3.abs\\
2b &Failed downregulation &gene1.pres, gene2.pres $\rightarrow$
gene3.pres\\
2c &No downregulation &gene1.pres, gene2.abs $\rightarrow$ gene3.pres\\
3 &Degradation &gene.pres $\rightarrow$ gene.abs\\
4 &No effect &gene.abs $\rightarrow$ gene.abs\\
\hline
\end{tabular}
\end{table}

Rules 3 and 4 are default rules; they are only applied to genes not occurring at the right side of
another rule.  Since the model abstracts from exact thresholds and time delays (which are rarely
known), there are the alternative downregulation rules 2a and 2b. After one time step, upregulation
or downregulation can prevail. (By the same reason, one could add to rule 3 the alternative
gene.present $\rightarrow$ gene.present.) The model is non-deterministic, the context $\mathcal{K}$
of Table \ref{transCon} shows the possible state transitions, starting from the initial state
$\varphi^{in}_0$ of the individual time series $\mathcal{K}^{obs}$. It could also be relevant to
investigate contexts containing the initial states of different observed cellular conditions,
different patients or with all possible input states.
\begin{table}\caption{\footnotesize Knowledge based transition context
$\mathcal{K}$ for 2
genes. Example rules for $(\varphi_1^{in},\varphi_2^{out})$:
Tnf$\alpha$.pres, Tnfaip3.pres $\rightarrow$ Tnf$\alpha$.abs (2a);
Tnf$\alpha$.pres,
Tnfaip3.pres $\rightarrow$ Tnfaip3.pres (2b).}\label{transCon}\centering
\begin{tabular}{cccccc}
\hline\noalign{\smallskip} \textbf{Transition} &\textbf{Tnf$\alpha^{in}$}
&\textbf{Tnfaip3$^{in}$}
&\textbf{Tnf$\alpha^{out}$} &\textbf{Tnfaip3$^{out}$} &\textbf{Applied rules}\\
\noalign{\smallskip} \hline \noalign{\smallskip}
$(\varphi_0^{in}, \varphi_1^{out})$ &+&-&+&+&2c\\
$(\varphi_1^{in}, \varphi_0^{out})$ &+&+&+&-&2b,2a\\
$(\varphi_1^{in}, \varphi_1^{out})$ &+&+&+&+&2b\\
$(\varphi_1^{in}, \varphi_2^{out})$ &+&+&-&+&2a,2b\\
$(\varphi_1^{in}, \varphi_3^{out})$ &+&+&-&-&2a\\
$(\varphi_2^{in}, \varphi_3^{out})$ &-&+&-&-&4,3\\
$(\varphi_3^{in}, \varphi_3^{out})$ &-&-&-&-&4\\
\hline
\end{tabular}
\end{table}

Implications of this context $\mathcal{K}$ simply reflect the rules applied in
order to compute a state
transition. Deterministic transition rules even may be included in the stem base
of the context,
or they follow from it. (Of course, the stem base contains also implications in the inverse
direction - from output to input attributes - or mixed implications like
gene1.pres$^{out}$,
gene2.abs$^{in} \rightarrow$ gene3.pres$^{in}$.)

The transitive context $\mathcal{K}_t$ is derived from $\mathcal{K}$ by adding all supplementary
objects $(\varphi^{in},\varphi^{out}) \in t(R)$. An interactive attribute exploration of
$\mathcal{K}$ may be more intuitive than an exploration of $\mathcal{K}_t$; the expert can compare
the implications in question to the measured one step transitions of $\mathcal{K}^{obs}$ and
eventually check them against supplementary knowledge. However, a time step of a knowledge based
transition is not identical to a measurement interval; the problem is aggravated, if the intervals
are different as in the present case. Therefore it seems more appropriate to explore the transitive
context $\mathcal{K}_t$ immediately. Its implications denote dependencies between attributes of
states related by transitions of arbitrary duration. The following procedure was applied:
\begin{enumerate}
\item Transform a time series of gene expression measurements to an observed
context $\mathcal{K}^{obs}$.
\item For a set of interesting
genes, extract transition rules from biological literature and
databases.
\item Construct the transition context $\mathcal{K}$, starting from
$\varphi^{in}_0$ of $\mathcal{K}^{obs}$.
\item Derive the respective transitive contexts $\mathcal{K}_t$
and $\mathcal{K}^{obs}_t$.
\item \label{compObs}Perform attribute exploration of $\mathcal{K}_t$. Decide
about an implication $A\rightarrow B$ by checking its validity in
$\mathcal{K}^{obs}_t$ and/or by searching for supplementary knowledge.
Possibly provide a counterexample from $\mathcal{K}^{obs}_t$.
\item Answer queries from the modified context $\mathcal{K}_t$ and from its
stem base.
\end{enumerate}

For all 5 genes Tnf$\alpha$, Tnfaip3, Icam1, Ccl4 and Il1$\beta$, a more
complex
set of transition rules had to be defined, which we shall not discuss
here.

In step \ref{compObs}, automatic decision criteria could be tresholds of support $q= |(A \cup B)'|$
and confidence $p= \frac{|(A \cup B)'|}{|A'|}$ for an implication in $\mathcal{K}^{obs}_t$. A weak
criterion is to reject only implications with support 0 (but if no object in $\mathcal{K}^{obs}_t$
has all attributes from A, the  implication is not violated). In the present example a strong
criterion was applied: implications of $\mathcal{K}_t$ had to be valid also in the observed
context. This is equivalent to an exploration of the union of the two contexts. Its results, where
all common implications were accepted by the expert, are presented in Table \ref{result}. It has to
be considered that the combined context generally only represents a transitive relation on the
states for its subcontexts $\mathcal{K}_t$ and $\mathcal{K}_t^{obs}$. The main purpose of the
proposed exploration is to make a falsification of the $\mathcal{K}_t$ implications possible.

The subsequent implications are noteworthy and biologically meaningful: 1. is equivalent to
always(Icam1.pres$^{out}$). The same assertion for Ccl4 was falsified by the measurement; instead
there are the new implications 2. to 5. and 15. to 17. The static implications 6. and 10. to 13.
reflect the very similar regulation of Il1$\beta$, Icam1 and Ccl4 (e.g. by Il1$\beta$ and
Tnf$\alpha$) and were also valid in the observed context. Likewise, 14. was supported by a priori
and observed transitions. 7. to 9. and 19. to 21. mirror the important role of the upregulating
genes Il1$\beta$ and Tnf$\alpha$: if Il1$\beta$, Tnfaip3 or Tnf$\alpha$ are upregulated at an
arbitrary time point, either Il1$\beta$ or Tnf$\alpha$ have been present in the past.
\begin{table}\caption{\footnotesize The stem base of the combined knowledge based and
observed transitive contexts. Implications following from the previously entered background
implications of the form gene.abs, gene.pres $\rightarrow \bot$ are not shown. The implications are
presented in a short form proposed during attribute exploration by the ConImp program (available at
http://www.mathematik.tu-darmstadt.de/$\sim$burmeister/ConImp.tar), with basis of premise and/or
reduced conclusion.}\label{result}\centering
\begin{tabular}{cll}
\hline\noalign{\smallskip} \textbf{Nr.} &\textbf{Implication}\\
\noalign{\smallskip} \hline \noalign{\smallskip}
1.& $\top$ &$\rightarrow$ Icam1.pres$^{out}$\\
2.& Il1b.abs$^{out}$ &$\rightarrow$ Ccl4.pres$^{out}$\\
3.& Ccl4.abs$^{out}$ &$\rightarrow$  Tnf$\alpha$.pres$^{in}$
Tnf$\alpha$.abs$^{out}$ Tnfaip.pr$^{out}$ Il1$\beta$.pres$^{out}$\\
4.& Tnfaip.abs$^{out}$ &$\rightarrow$ Ccl4.pres$^{out}$\\
5.& Tnf$\alpha$.pres$^{out}$ &$\rightarrow$   Ccl4.pres$^{out}$\\
6.& Il1$\beta$.pres$^{in}$ &$\rightarrow$  Icam1.pres$^{in}$ Ccl4.pres$^{in}$\\
7.& Il1$\beta$.abs$^{in}$  Il1$\beta$.pres$^{out}$  &$\rightarrow$
Tnf$\alpha$.pres$^{in}$\\
8.& Il1$\beta$.abs$^{in}$ Tnfaip.pres$^{out}$  &$\rightarrow$
Tnf$\alpha$.pres$^{in}$\\
9.& Il1$\beta$.abs$^{in}$ Tnf$\alpha$.pres$^{out}$  &$\rightarrow$
Tnf$\alpha$.pres$^{in}$\\
10.& Ccl4.pres$^{in}$ &$\rightarrow$ Icam1.pres$^{in}$\\
11.& Ccl4.abs$^{in}$ &$\rightarrow$  Tnf$\alpha$.pres$^{in}$ Tnfaip.abs$^{in}$
Icam1.abs$^{in}$ Il1$\beta$.abs$^{in}$\\
12.& Icam1.pres$^{in}$ &$\rightarrow$ Ccl4.pres$^{in}$\\
13.& Icam1.abs$^{in}$  &$\rightarrow$ Ccl4.abs$^{in}$\\
14.& Tnfaip.pres$^{in}$ &$\rightarrow$  Icam1.pres$^{in}$\\
15.& Tnfaip.abs$^{in}$ Icam1.pres$^{in}$ &$\rightarrow$   Ccl4.pres$^{out}$\\
16.& Icam1.pres$^{in}$ Ccl4.abs$^{out}$ &$\rightarrow$  Tnfaip.pres$^{in}$\\
17.& Tnfaip.abs$^{in}$ Ccl4.abs$^{out}$ &$\rightarrow$  Icam1.abs$^{in}$\\
18.& Tnf$\alpha$.abs$^{in}$ &$\rightarrow$  Icam1.pres$^{in}$
Ccl4.pres$^{out}$\\
19.& Tnf$\alpha$.abs$^{in}$ Il1$\beta$.pres$^{out}$ &$\rightarrow$
Il1$\beta$.pres$^{in}$\\
20.& Tnf$\alpha$.abs$^{in}$ Tnfaip.pres$^{out}\!\!$ &$\rightarrow$
Il1$\beta$.pres$^{in}$\\
21.& Tnf$\alpha$.abs$^{in}$ Tnf$\alpha$.pres$^{out}$ &$\rightarrow$
Il1$\beta$.pres$^{in}$\\
22.& Tnf$\alpha$.abs$^{in}$ Il1$\beta$.abs$^{in}$ &$\rightarrow$
Tnf$\alpha$.abs$^{out}$ Tnfaip.abs$^{out}$ Il1$\beta$.abs$^{out}$\\
\hline
\end{tabular}
\end{table}

By reasoning over the stem base, hypotheses and predictions as results of similar implicational
queries can be made, concerning transcriptome time series under equivalent experimental conditions
to those of $\mathcal{K}^{obs}$. A query B $\rightarrow$ eventually(m) (\ref{eventually}) is
decided positively for an existing transition path, if B $\rightarrow$ never(m) (\ref{never}) does
not follow from the stem base. Set operations in the resulting context provide answers to further
types of queries. It can be asked, whether a set of genes is in a stable state or shows an
oscillatory behaviour (\ref{oscillation}). Answers to queries such as (\ref{3point}) can explain an
observed 3-point time series.

Altogether experimental data can be
better understood, and reciprocally those are used for a validation of
the implicational knowledge base during the exploration process.

\section{Outlook}
A mathematically very interesting task will be the investigation of a new state
context; its objects are
states $\varphi$, and the attributes
are more abstract temporal properties like eventually(Ccl4.pres) or
oscillation($\varphi$). We want to develop a set of
background implications, so
that implications of the new context can be derived from those of the
transitive context. Also the dependency of a transitive from an underlying
transition context will be investigated. A continuous
task is to collect further meaningful biological
questions that can be answered by our approach, and to develop a
biologically more exact, comprehensive and realistic model. Thus it is
planned to introduce finer steps than present/absent and to adapt the transition
rules to this
approach. Also a more precise definition of time intervals could be
useful. Formal concept analysis is a mathematically and logically strict and
rich theory, and we
will further investigate its explanatory potential for temporal transitions.

\section{Acknowledgements}
I thank Bernhard Ganter / TU Dresden and Reinhard Guthke / Hans-Kn\"oll-
Institute Jena, for fruitful suggestions and discussions.

The work was supported
by the German Federal Ministry of Education and
Research BMBF (FKZ 0313652A).
\footnote{This paper has been published in G\'ely, A. et al.: \textit{Contributions to ICFCA 2007 - 5th
International Conference on Formal Concept Analysis}. Clermont-Ferrand 2007, 121-130.}

\nocite{GW99}


\begin{thebibliography}{1}
\bibitem{Cha04}
Chabrier-Rivier, N. et al.: Modeling and Querying Biomolecular Interaction
Networks. \textit{Theor. Comp.
Sc.} {\bfseries 325}(1) (2004), 25-44.

\bibitem{Glo06}
Glocker, M., Guthke, R., Kekow, J., Thiesen, H.-J.: Molecular Diagnostic
and Therapeutic Signatures of Rheumatoid Arthritis Identified by
Transcriptome and Proteome Analysis: On the Way Towards Personalized Medicine.
\textit{Medicinal Research Reviews} \textbf{26} (2006), 63-87.

\bibitem{GW99}
Ganter, B., Wille, R.: \textit{Formal Concept Analysis - Mathematical
Foundations}. Springer, Heidelberg
1999.

\bibitem{Rud01}
Ganter, B., Rudolph, S.: Formal Concept Analysis Methods for Dynamic Conceptual
Graphs. In: \textit{ICCS
2001}, LNAI 2120. Springer, Heidelberg 2001, 143-156.

\bibitem{Kau93}
Kauffman, S.A.: \textit{The Origins of Order: Self-Organization and Selection
in Evolution}. Oxford University Press, New York 1993.

\bibitem{KGC05}
King, R.D., Garrett, S.W, Coghill, G.M.: On the Use of Qualitative Reasoning to Simulate and
Identify Metabolic Pathways. \textit{Bioinformatics} {\bfseries 21}(9) (2005),
2017-2026.

\bibitem{Wol05}
Wolff, K.E.: States, Transitions, and Life Tracks in Temporal Concept Analysis.
In: Ganter, B., Stumme, S. and Wille, R.: \textit{Formal Concept Analysis -
Foundations
and Applications}, LNAI 3626. Springer, Heidelberg 2005, 127-148.
\end{thebibliography}
\end{document}